\documentstyle[preprint,aps]{revtex}

\begin{document}
\draft
\title{{\sl Ab initio} Investigation of Elasticity and Stability of Aluminium}
\author{Weixue Li \ \ \ \ \ Tzuchiang Wang}
\address{LNM, Institute of Mechanics, Chinese Academy of Science, \\
Beijing, 100080, China}
\maketitle

\begin{abstract}
On the basis of the pseudopotential plane-wave(PP-PW) method in combination
with the local-density-functional theory(LDFT), complete stress-strain
curves for the uniaxial loading and uniaxial deformation along the [001] and
[111] directions, and the biaxial proportional extension along [010] and
[001] of aluminium are obtained. During the uniaxial loading, certain
general behaviors of energy versus stretch and the load versus the stretch
are confirmed; in each acse, there exist three special unstressed
structures: f.c.c., b.c.c. and f.c.t. for [001]; f.c.c., s.c. and b.c.c. for
[111]. Using stability criteria, we find that all of these state are
unstable, and always occur together with shear instability, except the
natural f.c.c. structure. A Bain transformation from the stable f.c.c.
structure to the stable b.c.c. configuration cannot be obtained by uniaxial
compression along any equivalent [001] and [111] direction. The tensile
strength are similar for the two directions. For the higher energy barrier
of [111] direction, the compressive strength is greater than that for the
[001] direction. With increase in the ratio of the biaxial proportional
extension, the stress and tensile strength increase; however, the critical
strain does not change significantly. Our results add to the existing {\sl %
ab initio }database for use in fitting and testing interatomic potentials.
\end{abstract}


\preprint{}

\section{ Introduction}

Investigation of elastic behavior of perfect single crystal under loading is
of interest. It can, for example, be carried out on a system in which
substantial, elastic (but not necessary linear) deformation may occur, in
that case substantial deformation may occur either without significant
dislocation movement or before deformation caused by dislocation movement
become dominant. Deformation of whisker, twinning and martensitic
transformations are relevant examples. The instability and branching under
loading is related to ideal strength and transformation. Such information is
very useful in the analysis of structural response in solids, {\sl e.g.}
polymorphism, amorphization, and melting to fracture.

Although Born$^{\text{\cite{born}}}$ criteria are widely used in the
investigation of strength, they are only valid under zero load. On the basis
of a series of comprehensive theoretical and computational studies, Hill and
Milstein$^{\text{\cite{hill75}-\cite{mil82}}}$ pointed out the following:
(i): stability is relative and coordinate dependent;and (ii) different
choices of strain measure lead to different domains of stability. On the
basis of the Morse potential, they investigated the mechanical response of
perfect crystal, which included the stress-strain relation, instability,
branching and the strength of f.c.c. Cu$^{\text{\cite{mf80}}}$, f.c.c. Ni$^{%
\text{\cite{milstein73},\cite{mils80b},\cite{mils80c}}}$,and ${\alpha }$-Fe$%
^{\text{\cite{milstein71}}}$, for different loading modes ({\sl e.g}.
uniaxial loading, uniaxial deformation, and shear loading. here, `uniaxial
loading' means that uniaxial stress is applied to one axis, and the lateral
face is relaxed and stress-free(traction-free), while `uniaxial deformation'
just means that one axis is dilated or contracted, and the dimension of
other two axes are fixed; in uniaxial deformation, all three axes are
subjected to loading.) The loading direction which were adopted included
[001], [110] and [111]. Possible branching path were revealed. Using the
thermodynamics Gibbs function, Wang {\sl et al}$^{\text{\cite{wang93},\cite
{wang95}}}$ developed an equivalent-stability analysis method, and
investigated the onset of instability in a homogeneous lattice under
critical loading. The onset modes, derived from the stability criteria, were
verified by means of a molecular dynamics simulation. Zhou and Jo\'{o}s$^{%
\text{\cite{zhou96}}}$ derived general expressions for the stability
criteria by an appropriate thermodynamics potential.

There has been less investigation based on first principle. Senoo {\sl et al}%
$^{\text{\cite{senoo}}}$ discussed the elastic deformation due to [100]
loading of Al, using the pseudopotential method. Esposito {\sl et al}$^{%
\text{\cite{esposito}}}$ dealt with the tensile strength of f.c.c. Cu under
uniaxial deformation on the basis of the {\sl ab initio} potential,
augmented-spherical-wave(ASW), KKR methods. However, relaxation of the
crystal structure was not permitted. Paxton {\sl et al}$^{\text{\cite{paxton}%
}}$ calculated the theoretical strength of five b.c.c. transition metal, and
Ir, Cu, and Al by considering ideal-twin stresses using the full-potential
linear muffin-tin orbital (FP-LMTO). Sob {\sl et al.}$^{\text{\cite{sob}}}$
investigated the theoretical tensile stress in tungsten single crystal under
[001] and [111] loading by the FP-LMTO method. The stability analysis was
not explicit in any of the above cases.

Bain transformation takes a crystal from its stable b.c.c. configuration
into a stable f.c.c. structure, and vice versa, by means of homogeneous
axial deformations. Which path requires the lowest energy and stress barrier
between these states was investigated and reviewed by Milstein {\sl et al}$%
{\sl .}^{\text{\cite{mils94}}}$. The general mechanics and energetics of the
Bain transformation were presented. On the basis of the empirical
pseudopotential, Milstein {\sl et al}${\sl .}$$^{\text{\cite{mils94}}}$
investigated the Bain transformation of crystalline sodium in detail. These
kind of transformation is also relevant to the investigation of epitaxial
thin film$^{\text{\cite{fox}}}$.

In this paper, we present a direct investigation on the elasticity, the
stress-strain relation, the stability, and the ideal strength of f.c.c.
Aluminium within density functional theory. The stability analysis is
considered explicitly, on the basis of the theory of Hill, Milstein$^{\text{%
\cite{hill75}},\text{\cite{hill},\cite{mil82}}}$ and Wang {\sl et al.}$^{%
\text{\cite{wang93},\cite{wang95}}}$. We consider several loading modes:
uniaxial deformation and uniaxial loading along the [001] and [111]
directions, and biaxial proportional extension along [001] and [010]. The
deformation is homogeneous, elastic, and permitted to be appropriately
large. The stress-strain relation are calculated, and the ideal strength is
approached via the loss of stability. Branching or structure transformation
from a primary path of deformation takes place with the loss or exchange of
stability, which are relevant to the Bain transformation. In this way, the
mechanical responses for different loading modes and direction are obtained,
clearly from first principle,.

The paper is organized as follows. The calculation model is presented in
Sec. 2. In this section, we give the formulation of stress, elastic
stiffness coefficients, and stability criteria, especially for three loading
modes. Numerical precision is evaluated at the end of this section. As a
benchmark, equilibrium properties and elastic constants are calculated in
the Sec. 3, [001] uniaxial deformation and [001] uniaxial loading are
considered in Sec. 4, and a stability analysis is implemented. In Sec. 5,
[111] uniaxial deformation and [111] uniaxial loading are considered.
Results on the biaxial proportional extension are given in Sec. 6. {\sl Ab
initio} calculations can be used to construct a database for fitting and
testing interatomic potential$^{\text{\cite{rob94},\cite{payne96}}};$ a
brief discussion of our results together with the existing {\sl ab initio}
database for aluminium is given in Sec. 7. A summary and conclusions are
presented in the last section.

\section{ Formulation}

Consider an initial unstressed and unstrained configuration, denoted as $%
{\bf X}_{{\bf 0}}$. It undergoes homogeneous deformation under a uniform
applied load, and changes from ${\bf X}_{{\bf 0}}$ to ${\bf X=JX}_{{\bf 0}}$%
, where ${\bf J}$ is the deformation gradient or the Jacobian matrix and the
rotation part is subtracted. The associated Lagrangian strain tensor ${\bf E}
$ is: 
\begin{equation}
{\bf E}=\frac 12({\bf J}^{{\bf T}}{\bf J}-{\bf I})  \label{eq:eta}
\end{equation}
and the physical strain is: 
\begin{equation}
{\bf e}={\bf (J}^{{\bf T}}{\bf J)}^{\frac 12}-{\bf I}  \label{eq:ps}
\end{equation}

For the present deformation, internal energy {\sl U} is rotationally
invariant and therefore only a function of ${\bf E}${\bf .} The second
Piloa-Kirchhoff stress tensor ${\bf T}^{\text{\cite{true60}}}$ is defined
as: 
\begin{equation}
T_{ij}{\bf =}\frac 1{V_0}\frac{\partial {\sl U}}{\partial E_{ij}}
\label{eq:piloa}
\end{equation}
It relates Cauchy stress, {\sl i.e.}, true stress $\tau _{kl}$ , by the
following equation: 
\begin{equation}
T_{ij}=det|{\bf J}|{\sl J_{ik}^{-1}J_{jl}^{-1}\tau _{kl}}  \label{eq:tstress}
\end{equation}
where $det|{\bf J}|$ is the ratio $V/V_0$. With the Cauchy stress, the
applied force can be obtained by multiplying by the current transverse area.

For the stressed state ${\bf X}$, the elastic constants are determined
through the equation: 
\begin{equation}
C_{ijkl}({\bf X})=\frac 1{V({\bf X})}(\frac{\partial ^2{\sl U}}{\partial
E_{ij}^{^{\prime }}\partial E_{kl}^{^{\prime }}}\mid _{E^{^{\prime }}=0})
\label{eq:elast}
\end{equation}
where ${\bf E}^{\prime }$ is Lagrangian strain around the state ${\bf X}$.
These elastic constants are symmetric with respect to interchange of
indices, and are often expressed in condensed Voigt notation.

To analyse the stability, the elastic stiffness coefficient ${\bf B}^{\text{%
\cite{wang95}}}$ is introduced as follows: 
\begin{equation}
B_{ijkl}=C_{ijkl}+\frac 12(\delta _{ik}\tau _{jl}+\delta _{jk}\tau
_{il}+\delta _{il}\tau _{jk}+\delta _{jl}\tau _{ik}-2\delta _{kl}\tau _{ij})
\label{eq:stiff}
\end{equation}
From this definition, we can see that ${\bf B}$ does not posses $%
(ij)\longleftrightarrow (kl)$ symmetry generally. The system may be unstable
when 
\begin{equation}
det|{\bf B}|=0  \label{eq:stability1}
\end{equation}
for the first time.

The following loading modes are considered:

(i) Uniaxial Deformation. 
\begin{equation}
e_{ij}=e\delta _{i3}\delta _{j3}\ \ \ \ \ \ \ \ \ \ \ i,j=1,2,3
\label{eq:ud}
\end{equation}
In this mode, a strain is specified, and strain energy is evaluated by
subtracting a reference energy, which is calculated using the theoretical
lattice constant, on the basis of a total energy calculation. The
corresponding stretches of three axes are: $\lambda _1=\lambda _2=1$, $%
\lambda _3<1$ for compression, and $\lambda _3>1$ for tension.

(ii) Uniaxial Loading. 
\begin{equation}
\tau _{ij}=\sigma \delta _{i3}\delta _{j3}\ \ \ \ \ \ \   \label{eq:ul} \\
\ \ \  \\
\ \ i,j=1,2,3
\end{equation}
For a given longitudinal strain, let transverse lattice contract or dilate
to make the total energy arriving a minimum, which corresponds zero
stress(traction) on lateral faces. Due to the crystal symmetry, the
transverse contraction or dilation is same in the two transverse directions.
With the longitudinal strain and corresponding transverse strain, the strain
energy of the uniaxial loading is calculated. The corresponding stretches of
three axes are: for compression, $\lambda _1=\lambda _2>1$, and $\lambda
_3<1 $; for tension, $\lambda _1=\lambda _2<1$, and $\lambda _3>1$.

(iii) Biaxial Proportional Extension 
\begin{equation}
e_{22}=\alpha e_{33}\not{=}0\ \ \ \ e_{ij}=0\ \ \ otherwise  \label{10}
\end{equation}

{\sl Ab initio} pseudopotential plane-wave method is implemented. On the
basis of the mechanism of Hamman $^{\text{\cite{hamman}}}$ and Troullier and
Martins$^{\text{\cite{troullier}}}$, soft first principle pseudopotentials
are generated using the package DgncppB$^{\text{\cite{fuchs},\cite{gonze}}}$%
. The package Fhi96md$^{\text{\cite{fhi96md}}}$ , which employs a
first-principle pseudopotential and a plane-wave basis set, is used to
perform the DFT total-energy calculation. In our calculations, local density
approximation(LDA) with the exchange and correlation energy functional
developed by Perdew and Zunger$^{\text{\cite{perdew}}}$ is adopted.

Two supercell are designed in our calculations: one is four-atom supercell,
for use in considering the equilibrium properties and elastic constants,
uniaxial deformation and loading along [001], and biaxial deformation along
[010] and [001]; the other one is three-atom supercell for use in
considering uniaxial deformation and loading along [111]. Due to the
calculation of stress and elastic constants, the precision must be evaluated
carefully. $E_{cut}$=12Ry is sufficient in our pseudopotential plane-wave
calculations. Due to discontinuous feature of occupation number of metal
electron, a large number of k-space samples must be used to reach sufficient
precision. A smear parameter $\Delta $ is introduced to decrease the number
of k-points. Our calculation show that $\Delta =0.058Ry$ already gives a
satisfactory result. The corresponding k-meshes are $8\times 8\times 8$ for
the four-atom supercell, and $10\times 10\times 6$ for three-atom supercell.

\section{ Equilibrium Properties and Elastic Constants}

As a test, we have calculated the equilibrium lattice constant and elastic
constants of bulk Al. Face-centered-cubic Al has three independent elastic
constant, {\sl i.e}${\sl .}$ $C_{11}$, $C_{12}$, $C_{44}$. The equilibrium
lattice constants $a_0$ the bulk modulus ${\sl B}_0$ are obtained by fitting
energy-volume curve to Murnaghan equation of state.$^{\text{\cite{murnaghan}}%
}$ The relation between bulk modulus ${\sl B}_0$ and elastic constant is $%
B_0=(C_{11}+2C_{12})/3$. From the uniaxial deformation along [001] direction
and trigonal strain along [111] direction, $C_{11}$ and $C_{44}$ are
obtained. The result are given in table 1.

The theoretical lattice constant, $a_0=3.97\AA $, is 2$\%$ smaller than
experimental value $a_0=4.05\AA $, and the corresponding elastic constants
are 10$\%$ larger than experimental values$^{\text{\cite{simmons}}}$
measured at room temperature. This difference is typical of DFT-LDA
calculations, and can be see clearly from other first principle results. Sun 
{\sl et al.}$^{\text{\cite{sun}}}$ pointed out the elastic constants are
sensitive to the lattice constant of the crystal, and calculated the
corresponding data using the lattice constants at room temperature. On the
basis of the linear augmented plane wave(LAPW) method, Mehl and Klein$^{%
\text{\cite{mehl}}}$ made similar calculations, and gave the Bulk modulus,
Young's modulus, shear modulus, average Possion ratio and anisotropy of
isotropic materials with an orientation average. We also calculated the
elastic moduli using the experimental lattice constants, we present the
results in table 1. All of the theoretical calculations are in good
agreement with experimental data, and better result are given by our
calculation. However, it is worthy noting that we performed our calculations
of the strength and stability using the theoretical lattice constant. Any
externally imposed strain and stress should be excluded to get accurate
results.

\section{ [001] Uniaxial Loading and [001] Uniaxial Deformation}

A four-atom face-centred tetragonal supercell was designed for investigating
[001] uniaxial loading and [001] uniaxial deformation. [100], [010] and
[001] is selected as X-, Y- and Z-axis. In the initial equilibrium state,
the f.c.t. structure is f.c.c. With the same f.c.t. cell, b.c.t. cell can be
got by means of an axial rotation by $\frac \pi 4$ around [001] axial. This
relation can be found from figure 1. We start from the unstressed f.c.c.
state where $\lambda _1=\lambda _2=\lambda _3=1$, and let the Z-axis be
compressed: on the prescribed path, the lattice must pass through the state $%
\lambda _3=\lambda _1/\sqrt{2}=\lambda _2/\sqrt{2}$ where the b.c.t. becomes
a b.c.c. one. The cubic symmetry at this point implies that the loads are
hydrostatic. For [001] uniaxial loading, in particular, since the transverse
loads are always zero, the axial load at this state must be zero. Since the
load must be tensile as $\lambda _3\rightarrow \infty $ and compressive as $%
\lambda _3\rightarrow 0$, the existence of two unstressed states on the
primary path of [001] loading implies a third zero point in general.
Obviously, the third one does not have any symmetry higher than tetragonal.
The state located in the central unstressed position is always unstable at
its local energy maximum. For a detailed analysis and proof of the general
form of the energy versus stretch and stress versus stretch, the reader is
referred to the original paper of Milstein$^{\cite{mils80a},\cite{mils80e}}$.

For tetragonal symmetry of the crystal under uniaxial loading and uniaxial
deformation along the [001] direction, the number of independent elastic
constants is reduced to six: $C_{33}$, $C_{12}$, $C_{13}=C_{23}$, $%
C_{11}=C_{22}$, $C_{44}=C_{55}$ and $C_{66}$; all of the other $C_{ij}$ are
equal to zero. Since we are more interested in uniaxial loading, we analyse
its stability. With Eq. \ref{eq:stiff}, \ref{eq:stability1} and \ref{eq:ul},
we write the instability criteria of [001] uniaxial loading as follows: 
\begin{eqnarray}
(C_{33}+\sigma )(C_{11}+C_{12})-2C_{13}(C_{13}-\sigma ) &\leq &0
\label{eq:uls} \\
C_{11}-C_{12} &\leq &0 \\
C_{44}+\frac 12\sigma &\leq &0 \\
C_{66} &\leq &0
\end{eqnarray}
The first expression involves the vanishing of the bulk modulus, and is
referred to as the spinodal instability criterion. The second instability
involves symmetry breaking (bifurcation) with volume conservation; it may be
identified as the tetragonal shear breaking, and is referred to as Born
instability. In this case, the crystal can branch from the tetragonal path
to a face-centred orthorhombic path under uniaxial dead loading; that is,
the branching is $\delta \lambda _1=-\delta \lambda _2\neq 0$ with $\delta
\lambda _3=0$ and $\delta \tau _{11}=\delta \tau _{22}=\delta \tau _{33}=0$.
The condition $C_{66}=0$, when the $C_{ij}$ are reckoned relative to the
f.c.t. crystal axes, is equivalent to $C_{11}-C_{12}=0$ when the $C_{ij}$
are computed relative to the axes of the b.c.t.cell. So, for the state where 
$C_{66}=0$(referred to the f.c.t. axes), the tetragonal crystal branch to a
body-centred orthorhombic path under uniaxial loading. $C_{44}+\frac 12%
\sigma =0$ gives another shear instability.

Energy versus strain and stress, force versus strain curve are given in
figure 2 (unless otherwise stated, the strain, force and stress, given in
figures are the physical strain, applied force, and Cauchy stress.) There
exist only one energy minimum under uniaxial deformation. Any departure from
the minimum for this loading mode lead to a rapid increase of the strain
energy. Due to the triaxial stresses for the uniaxial deformation, its
strain energy is always greater than that for the uniaxial loading. This is
same as the Milstein's conclusion: the uniaxial loading represents the
lowest-energy path between any two Bain paths. However, another local
maximum and local minimum state are found under uniaxial compression. The
longitudinal and transverse strain at the local maximum are $e_{33}=-0.20$
and $e_{11}=0.1313$ respectively; the ratio of stretch $%
(1-0.2)/(1+0.1313)=0.7071$ , is approximately equal to $1/\sqrt{2}$. The
corresponding structure is b.c.c. with lattice constant $3.176\AA $, as
expected. The remaining local minimum is a f.c.t. structure with $%
e_{33}=-0.305$. From figure2(b), we see that all three extreme are
stress-free. Because b.c.c. is located at a local maximum, it is unstable.

The elastic constants and corresponding stability range under [001] uniaxial
loading are shown in figure 3. $C_{66}$ is always negative over the range
[-0.40, -0.128], which includes the unstressed f.c.t. state. That means
that, although the stress-free f.c.t. is at the local minimum of the
uniaxial loading , it is still unstable against shear loading. On the basis
of the explanation of $C_{66}$, we wee that this f.c.t. state can transform
to a body-centred orthorhombic under uniaxial compression. The range of
spinodal instability under compression is [-0.263, -0.119]. The b.c.c. state
lies at a double instability.

From figure 2(b), the compressive strength of the [001] uniaxial loading is
-5.62 GPa with $e_{33}=-0.1$. During tension, the stress approaches its
maximum 12.54 GPa with $e_{33}=0.36$. However, $C_{11}-C_{12}$ and $C_{44}+%
\frac 12\sigma $ have already become negative when $e_{33}$ exceeded $0.272$%
, and the corresponding stress is 12.1 GPa. This gives the lower limit of
the tensile strength. Two possible branching are triggered at this critical
strain. With $C_{11}-C_{12}=0$, tetragonal lattice will transform via a
face-centred orthorhombic path and finally become a stable b.c.c. state at
last$^{\text{\cite{mils80d}}}$. With $C_{44}+\frac 12\sigma =0$, the
orthorhombic symmetry will be lost. However, which branching takes place
depends on the higher-order elastic modulus. By means of pseudopotential
methods based on a proposed model potential, Senoo {\sl et al.}$^{\text{\cite
{senoo}}}$ obtained a compressive strength of approximately -5.0 GPa, with
strain -0.11, and unstressed b.c.c. structure occurred at a strain of -0.2.
These results are very similar to ours. However, the tensile strength, which
they obtained, 17.4 GPa with a strain of 0.42, is greater than ours. The $%
C_{11}-C_{12}=0$ branching was assumed to take place at a strain 0.15 in
their work. Paxton {\sl et al.} $^{\text{\cite{paxton}}}$ calculated the
ideal-twin stress of Al on the basis of the FP-LMTO method. The
corresponding value, $0.14\times \frac 13(C_{11}-C_{12}+C_{44})=4.61$ GPa
(here the $C_{ij}$ are the elastic constants of theoretical lattice), is
approximately one third of our tensile strength.

Unlike the findings for Ni$^{\text{\cite{milstein73},\cite{mils80b},\cite
{mils80c}}}$ and Cu$^{\text{\cite{mf80}}}$, ${\alpha }$-Fe$^{\text{\cite
{milstein71}}}$ calculated by Milstein {\sl et al.}, the uniaxial stress and
force of Al are always lower than those for uniaxial deformation before they
approach the maximum. The transverse strain versus longitudinal strain is
given in figure 4, and the corresponding Possion ratio is positive along the
whole path of uniaxial loading.

\section{ [111] Uniaxial Loading and [111] Uniaxial Deformation}

For the case of [111] uniaxial loading and [111] uniaxial deformation, the
supercell is designed as follows: the planar vectors are identical to two
primitive f.c.c. lattice vectors, for instance along [110] and [101]
directions; the third lattice vectors is in the [111] direction; and here we
select three layers. In each layer, there is only one atom.

The path of deformation considered is axisymmetric, and all directions
transverse to [111] are equally stretched or fixed. The axisymmetric path of
deformation under [111] loading and [111] deformation consequently passes
through three cubic configurations: f.c.c., s.c. and b.c.c. with increase of
the compression. To illustrate this, a single quantity, r, the ratio of the
longitudinal to the transverse stretch under loading, is defined. Three
smallest tetrahedra are cut separately from f.c.c., s.c., and b.c.c. as
shown in figure 5. The bottom plane ABC is just the $\{111\}$ plane, and the
direction of OD is [111]. The ratios of height and edge length of bottom
plane are $\frac{\sqrt{6}}3$, $\frac{\sqrt{6}}6$, $\frac{\sqrt{6}}{12}$, 
{\sl i.e.} 1, 0.5, 0.25. These ratios are just values of r, which we define
above. When r decreases from 1 to 0.25 during compression, the cubic f.c.c.,
s.c. and b.c.c. structures appear in turn. The details can be found from
figure 5. During [111] uniaxial loading, the transverse load is always zero,
and cubic symmetry requires three cubic configurations to be stress-free.
However, during [111] uniaxial deformation, there is always exist a
transverse load, and cubic symmetry of s.c. and b.c.c. indicates hydrostatic
compression. For the transverse contraction under [111] loading, the s.c.
and b.c.c. states will occur earlier than the [111] deformation.

The calculated results are given in figure 6. Just like in the [001] case,
there exist only one minimum under uniaxial deformation, and the strain
energy is always higher than the uniaxial loading. Under uniaxial loading,
another local maximum and local minimum are obtained; the form of energy
versus stretch relation is just same as for [001] loading. From figure 4, we
get $e_{33}=-0.333$ at the local maximum point with $e_{11}=-0.334$; the
corresponding r, 0.5, is simply that for the unstressed s.c. configuration.
This state is unstable because it is located at a local energy maximum. At
the local minimum, $e_{33}=-0.59$ and $e_{11}=0.64$; the corresponding r,
0.25, is that for unstressed b.c.c. structure. From a simple geometric
calculation, we obtain the lattice constant of stress-free s.c. and b.c.c.
structures: $2.648\AA $ and $3.253\AA $ respectively. The corresponding
elastic constants are: (s.c.) $C_{11}+C_{12}=-10.1$ GPa, $%
C_{11}-C_{12}=-24.9 $ GPa and $C_{44}=4.4$ GPa; (b.c.c.) $%
C_{11}+C_{12}=65.19 $ GPa, and $C_{11}-C_{12}=-48.92$ GPa and $C_{44}=26.7$
GPa. From these values, we see that both of s.c. and b.c.c. structures are
unstable --- even the b.c.c. structure located at the local minimum of the
uniaxial loading. A shear instability always accompanies the unstressed s.c.
and b.c.c. state. On the basis of present and previous discussions, we
conclude that for f.c.c. Al, a stable b.c.c. structure cannot be obtained by
uniaxial compression along any equivalent [001] and [111] directions. The
possible Bain transformation from a stable f.c.c. to a stable b.c.c. is a
branching caused by uniaxial tension.

Because of the lower symmetry under [111] loading and numerical feature of
first principle calculations, the analysis of stability under this loading
mode is difficult, and is hence omitted except for the several special
points, {\sl i.e.} the initial f.c.c., unstressed s.c. and b.c.c. states.
From figure 6(b), we can see that the stress and force for the uniaxial
loading are always smaller than those for the uniaxial deformation. The
maximum of the tensile stress is 11.05 GPa at $e_{33}=0.295$, and the
maximum magnitude of compressive stress is 15.89 GPa at $e_{33}=-0.177$. The
tensile strength and critical strain are similar to those for [001] tension;
however, the compressive strength and critical strain are significantly
great than those for the [001] uniaxial compression. (The details are given
in table 2.) This can be attributed to the symmetry of the materials. Under
uniaxial compression, the stable f.c.c. crystal in the [001] case can
transform to another two extreme point more easily than in the [111] case.
This points is obvious from the following comparison. Under [001] loading,
the stress-free b.c.c. and f.c.t. are approached at $e_{33}=-0.2$ and $%
e_{33}=-0.305$, where the f.c.t. structure represents the local minimum and
the initial f.c.c. structure represent the overall minimum. The energy
barrier from these minimums is as follows: $\Delta E_{f.c.c.\rightarrow
f.c.t.}=0.1047$(ev/atom), $\Delta E_{f.c.t.\rightarrow f.c.c.}=0.0319$%
(ev/atom). For the [111] loading, the strain of the unstressed s.c. and
b.c.c. configuration are -0.333 and -0.59, where the b.c.c. structure
represents the local minimum. The corresponding energy barrier is as
follows: $\Delta E_{f.c.c.\rightarrow b.c.c.}=0.3766$(ev/atom), $\Delta
E_{b.c.c.\rightarrow f.c.c.}=0.2766$(ev/atom). A higher energy barrier to
transition and a higher critical strain are needed for [111] compression.

\section{ Biaxial Proportional Extension}

Biaxial proportional extension is considered here. In the present paper, we
deal with extension along the [010] and [001] directions; volume relaxation
are not considered. The strain ratio between [010] and [001] are 0.25, 0.5
0.75 and 1. The result is given in figure 7. With increase of the ratio, the
energy, stress and maximum stress increase, respectively. This is because
more energy is needed with a higher transverse strain for the same
longitudinal strain. However, the critical strains are similar for different
proportional loading modes.

\section{{\sl Ab initio} database of Aluminium}

Using total-energy calculations from first principle, Robertson {\sl et al.}$%
^{\text{\cite{rob94}}}$ and Payne {\sl et al.}$^{\text{\cite{payne96}}}$
constructed an {\sl ab initio} database for aluminium, which includes 171
structures with coordination number ranging from 0 to 12 and
nearest-neighbor distance ranging from 2.0 $\AA $ to 5.7 $\AA $, for fitting
and testing interatomic potentials. The energies (per atom) of all
structures are listed with the corresponding nearest-neighbor distances. The
basic 18 structures are calculated self-consistently, and the remaining 153
structures, obtained using non-self-consistent calculations, were generated
from the hydrostatic pressure. However, force, range of stability and
branching points, which are important for determining the topology of energy
surface, were not given in \cite{rob94} and \cite{payne96}.

In our investigations, all of the total energy calculations are
self-consistent; the basic structure is four-atoms and three-atom f.c.c.
supercells; three loading modes({\sl \ i.e.} uniaxial deformation, uniaxial
loading and biaxial proportional extension) and two loading directions ( 
{\sl i.e. }[001] and [111] ) are considered. The uniaxial loading path
connects structures with different coordination numbers, for example, for
the [111] case, from right to left, 12 (f.c.c.)$\rightarrow $6 (s.c.)$%
\rightarrow $8 (b.c.c.), and for the [001] case, 12 (f.c.c.)$\rightarrow $8
(b.c.c.). The complete energy and force, stress curve are plotted. In
particular, the ranges of stability and branching points of [001] uniaxial
loading are presented. The stabilities of three extreme points of [111]
uniaxial loading are also given. These information is essential to the
determination of functional form of the interatomic potential.

\section{ Summary and Conclusion}

On the basis of DFT total-energy calculations and stability theory given by
Hill and Milstein$^{\text{\cite{hill75}},\text{\cite{hill},\cite{mil82}}}$
and Wang {\sl et al.}$^{\text{\cite{wang93},\cite{wang95}}}$, we have given
a detail investigation of the mechanical response of f.c.c. aluminium for
different loading modes and loading directions. We reached the following
conclusions:

(1). In view of the requirement of crystal symmetry, the general form of the
energy versus stretch, and stress versus stretch relations for uniaxial
loading of cubic crystal can be described as follows: one local minimum, one
local maximum, and one overall minimum for energy versus stretch curve,
which are related with three unstressed state. The state in the middle is
always unstable because of its positioning at an energy maximum.

However, when we consider the a complicated crystal, for example one with
diamond structure, or alloys$^{\text{\cite{cral94},\cite{sob2}}}$, the
conclusions above should be treated with caution. In such case, the symmetry
is determined by both the structural parameters and the atomic ordering, and
some symmetry-dictated extrema may be lost.

(2). The complete stress-strain curves for uniaxial deformation and uniaxial
loading along the [001] and [111] direction, and for biaxial proportional
extension are obtained. The magnitudes of the stress and force for the
uniaxial deformation is always greater than those for the uniaxial loading
in the [001] and [111] directions over the range of stability. The stability
range for [001] uniaxial loading is given explicitly. The tensile and
compressive strength along the [001] and [111] directions are presented.

(3). Along the path of [001] uniaxial loading, the local minimum, local
maximum and overall minimum correspond to unstressed f.c.t., b.c.c. and
f.c.c. structures; under [111] uniaxial loading, they relate to stress-free
state b.c.c., s.c. and f.c.c. respectively. The intermediate b.c.c. for
[001] uniaxial loading and s.c. for [111] uniaxial loading are unstable as
they lie at local maximum. Although the f.c.t. state for [001] uniaxial
loading and b.c.c. state for uniaxial [111] loading lie at the local
minimum, they are still unstable against the shear instability. It is worthy
noting that the b.c.c. configuration for the [001] and [111] loading are
different configuration. Their lattice constants are 3.176$\AA $ and 3.253$%
\AA $, and the former is located at a local maximum while the later is
located at a local minimum.

A stable b.c.c. state of Al metal cannot be obtained by uniaxial compression
along any equivalent [001] and [111] direction. The possible Bain
transformation is a branching from the prescribed path of uniaxial tension
along equivalent [001] or [110] directions accompanied by branching.

(4). The tensile strength is similar along the [001] and [111] directions.
For the higher energy barrier for [111] uniaxial compression, the
compressive strength is greater than in the [001] case.

(5).The present results add to the existing {\sl ab initio} database.

\acknowledgements

This work was supported by the National Natural Science Foundation of China
(Grant No.19704100) and the National Natural Science Foundation of Chinese
Academy of Science (Grant No. KJ951-1-201). One of authors, Weixue Li,
thanks Prof. D S Wang for a useful discussion and encouragement. Weixue Li
also thanks Dr. Y. Yao for transferring package DgncppB to the Linux system.
Parts of computations are performed on the super-parallel computer of the
Network Information Center of the Chinese Academy of Science.

\begin{table}[tbp]
\caption{The strength of Al calculated for different loading modes; here, TS
means tensile strength, and CS means compressive strength.}
\begin{center}
\begin{tabular}{c|cccc}
& Uni. Deformation & $\epsilon_{33} \, \ \epsilon_{11}=\epsilon_{22}$ & Uni.
Loading & $\epsilon_{33} \, \ \epsilon_{11}=\epsilon_{22}$ \\ \hline
$[001$ TS & 12.65 (GPa) & 0.30, 0. & 12.1 (GPa) & 0.272, -0.0386 \\ 
$[001]$ CS &  &  & -5.62 (GPa) & -0.1, 0.049 \\ 
$[111]$ TS & 11.52 (GPa) & 0.265,0. & 11.05 (GPa) & 0.295, -0.0453 \\ 
$[111]$ CS &  &  & -15.89 (GPa) & -0.177, 0.110 \\ 
Shear strength $^{\cite{paxton}}$ & 4.61 (GPa) &  &  & 
\end{tabular}
\end{center}
\end{table}

\newpage

\begin{figure}[tbp]
\caption{Two fundamental cells of the face-centred tetragonal lattice and
body-centred tetragonal lattice within the same lattice}
\end{figure}

\begin{figure}[tbp]
\caption{The calculated strain energy (a), force and stress (b) during [001]
deformation and [001] loading for the theoretical lattice constant.}
\end{figure}

\begin{figure}[tbp]
\caption{The calculated elastic constants (a) and stability (b) undering
[001] loading for the theoretical lattice constant.}
\end{figure}

\begin{figure}[tbp]
\caption{The calculated transverse strain for [001] loading and [111]
loading with the theoretical lattice constant.}
\end{figure}

\begin{figure}[tbp]
\caption{The three smallest tetrahedra cut separately from f.c.c., s.c., and
b.c.c. structures. The length unit of each tetrahedron is its lattice
constant of the corresponding original lattice. The bottom plane is just the 
$\{111\}$ plane, and the triangle ABC is an equilateral triangle; the
direction of OD is [111]. The ratio of OD to CB is the ratio of longitude
stretch to transverse strecht.}
\end{figure}

\begin{figure}[tbp]
\caption{The calculated strain energy (a), force and stress (b) for [111]
deformation and [111] loading with the theoretical lattice constant.}
\end{figure}

\begin{figure}[tbp]
\caption{The calculated strain energy(a) and stress(b) during biaxial
proportional extension, with diffenerce ratio along the [010] and [001]
directions, for the theoretical lattice constant.}
\end{figure}

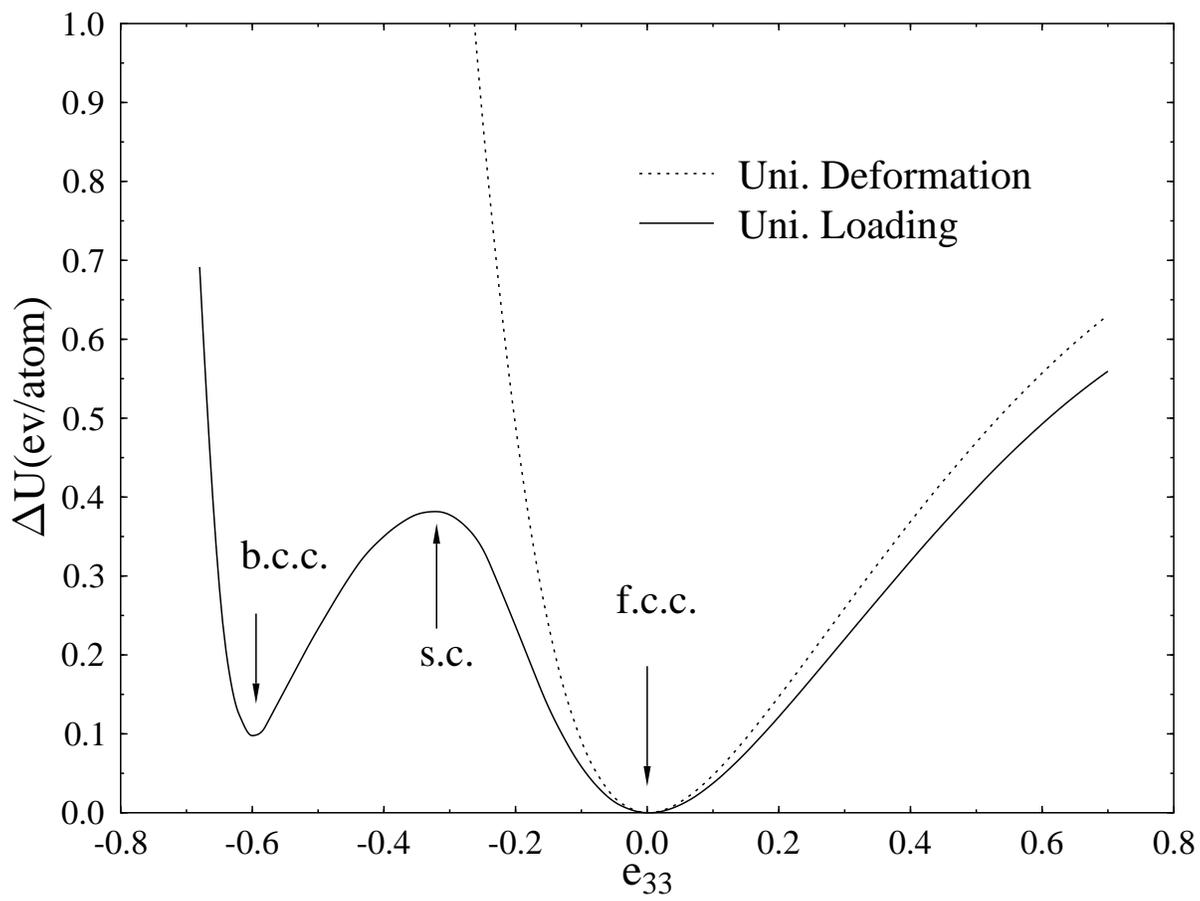
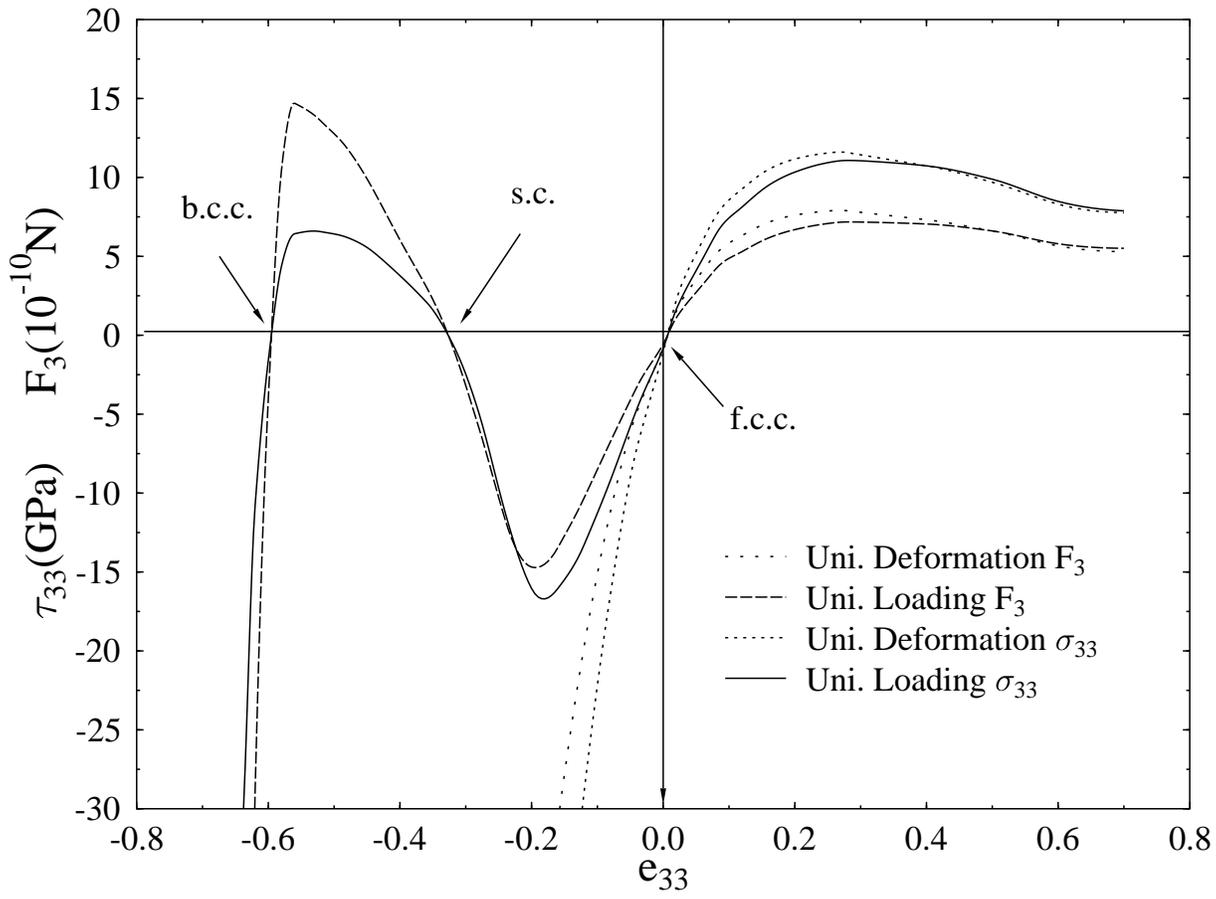
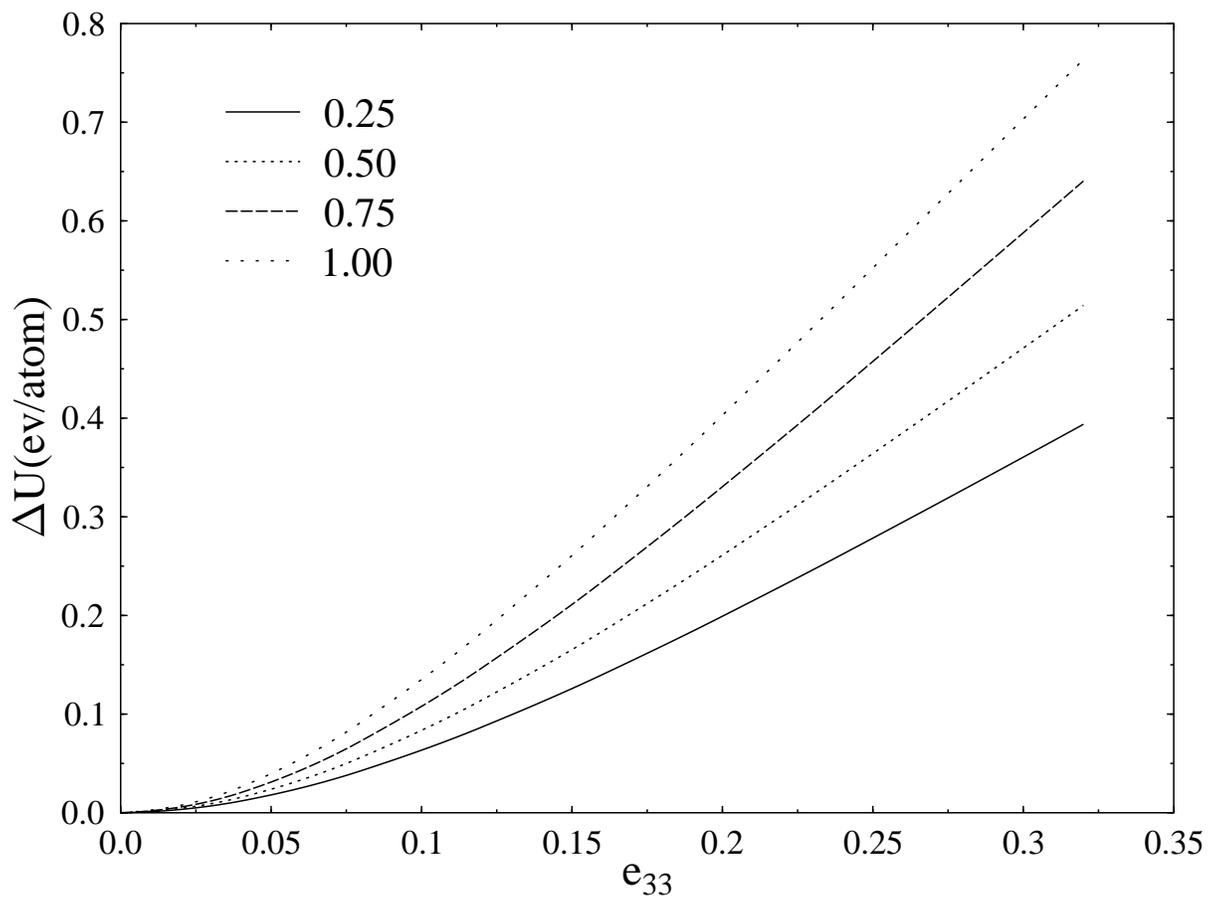
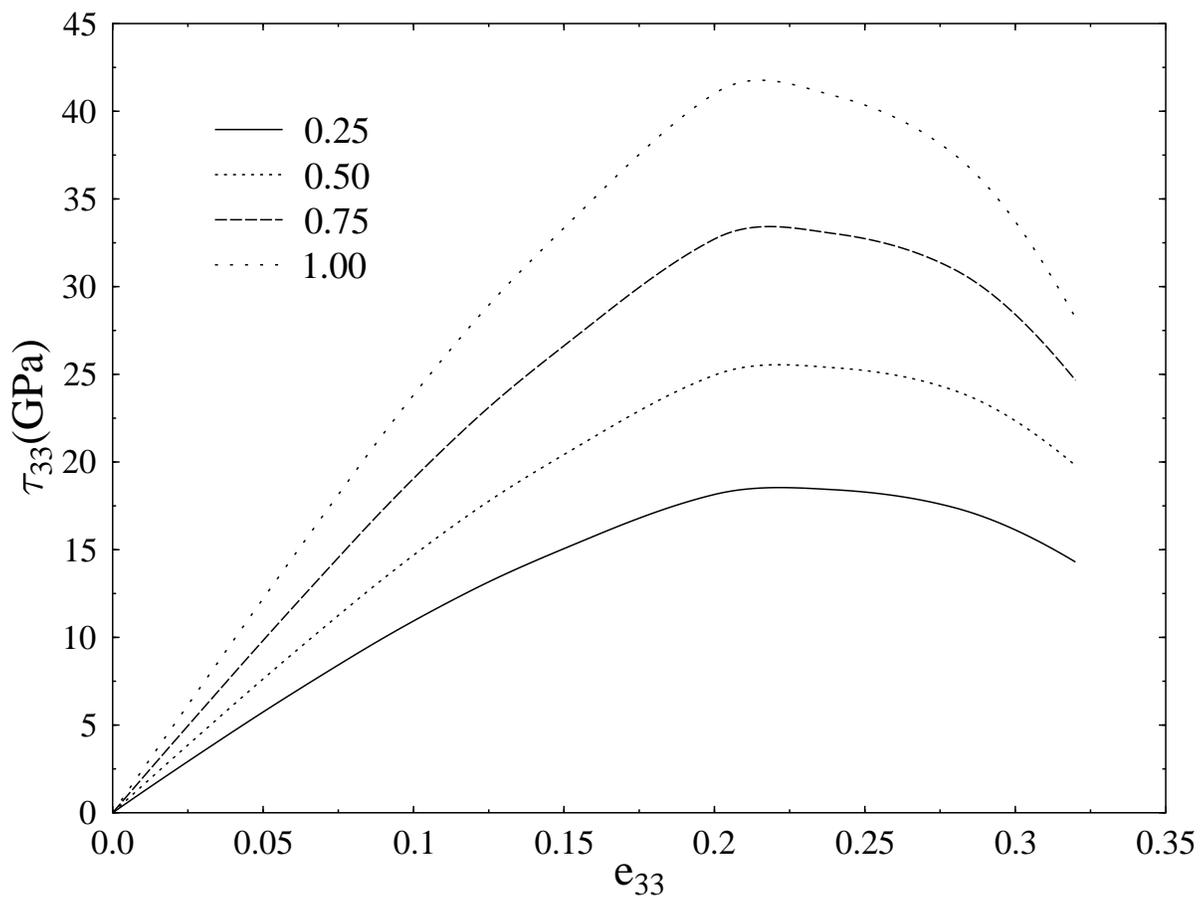
\end{document}